\documentclass[aps,twocolumn,10pt]{revtex4}
\usepackage{amsfonts}
\usepackage{amsmath}
\usepackage{amssymb}
\usepackage{graphicx}

\input{tcilatex}

\begin{document}

\title[Short title for running header]{Alkali Gases in Optical Lattices: A
New Type of Quantum Crystals?}
\author{A. E. Meyerovich}
\affiliation{Department of Physics, University of Rhode Island, Kingston, RI 02881 - 0817}

\begin{abstract}
Similarities between alkali gases in optical lattices with non-integer
occupation of the lattice sites and quantum crystals are explored. The
analogy with the vacancy liquid (VL) provides an alternative explanation to
the Mott transition for the recent experiment on the phase transition in the
lattice. The VL can undergo BEC with $T_{c}$ within experimental reach.
Direct and vacancy-assisted mechanisms of the band motion for hyperfine
impurities are discussed. Large concentration of vacancies can result in the
spatial decomposition of the system into pure hyperfine components. Below
the vacancy condensation the impurity component resembles $^{3}He$ in $%
^{3}He-HeII$ mixtures.
\end{abstract}

\volumeyear{}
\volumenumber{}
\issuenumber{}
\eid{}
\startpage{1}
\endpage{}
\maketitle

Recently, after the spectacular experimental discovery of Bose condensation,
the study of alkali gases in traps has become the focal point in atomic, low
temperature, and condensed matter physics. One of the most fascinating
features is the possibility to see in an experiment some of phenomena that
have been discussed earlier only within theoretical models (see review \cite%
{cor1}). An additional attraction is that the phenomena in ultracold alkali
gases are incredibly rich and combine features inherent to diverse condense
matter and low temperature systems (Refs. \cite{review2} and references
therein). For example, BEC condensation in trapped gases resembles, but is
not quite the same as the transition in other superfluid or superconducting
systems \cite{review2}. Another example is the dynamics of the hyperfine
components which resembles the spin dynamics of spin-polarized quantum gases 
\cite{jila1}.

A new example is an ultracold alkali gas in an optical lattice. Alkali atoms
are almost localized in microscopically periodic potential wells induced by
the Stark effect of interfering laser beams (Refs. \cite{lat1} and
references therein). The tunneling probability $t$ between the wells is
determined by the depth and size of the wells, \emph{i.e.}, by the intensity
and the wavelength $\lambda $ of the beams. Since this intensity is
adjustable, the atoms can be studied in a wide range of the tunneling
frequencies $t$ and effective masses $m^{\ast }\sim \hbar ^{2}/ta^{2}$, from
an almost free gas with a periodic perturbation to a well-localized "solid" (%
$a=\lambda /2=\pi /k$ is the lattice period). Another parameter is the
on-site repulsion $U$ for atoms inside the same well.

The standard Hubbard model for electrons predicts \cite{mott1} that the
alkali atoms in the optical lattice should exhibit\ an analog of the Mott
metal-insulator transition at $U/t\approx 5.8z$ ($z$ is the number of the
nearest neighbors). At $U\gg t$, the system should become an "insulator"
without interwell transitions (the particle tunneling between the lattices
site increases the on-site energy by $U$ and is energetically prohibitive).
At $U\ll t$, the on-site interaction does not restrict tunneling between the
sites and the atoms are in the "metal" phase. Then, at sufficiently low
temperature, the system can undergo BEC into a lattice superfluid. The
energy parameters $U$, $t$, and the potential well $V_{0}$ are often
measured in units of the recoil energy $E_{r}=\hbar ^{2}k^{2}/2m$. A typical
example is \cite{mott1} $t/E_{r}\sim 0.07$ and $U/E_{r}\sim 0.15$ for $%
V_{0}/E_{r}=15$.

This type of transition has been reported in Ref. \cite{nature1} for
lattices of the size $a=\pi /k\sim 426$ $\mathrm{nm}$ with recoil energy $%
E_{r}=\hbar ^{2}k^{2}/2m\sim 1$ $\mathrm{kHz}$. At low beam intensity,\emph{%
\ i.e.,} at large $t$, the experiment revealed the condensate peak in the
center of the trap. This peak disappeared at small $t$ (at $V_{0}/E_{r}$
between 13 and 22) which might indicate the transition to the Mott insulator
(MI) phase. However, the identification of the high-$U$ phase as the MI is
not unambiguous. The MI can be observed only when the average number of
atoms on the same lattice site is integer. If the average population is
fractional, the highest on-site energy states are not fully occupied. The
tunneling of the "excessive" particles from the site on which the highest
level is occupied to the site with an unoccupied level cannot be banned by
the on-site interaction. The tunneling of the "excessive" particle from site 
$\mathbf{r}_{1}$ to the empty neighboring site $\mathbf{r}_{2}$\ increases
the on-site energy by $U$ on the site $\mathbf{r}_{2}$\ while simultaneously
decreasing it by $U$ on the vacated site $\mathbf{r}_{1}$. Since both sites
are translationally equivalent, this opens the way to the band motion of the
"excessive" particles and to the existence of the partially filled
conduction band. Then the lattice with non-integer occupation stays in the
"metal" or "semiconductor" state even at large $U$ with the "excessive"
particles in the conduction band. Ref. \cite{nature1} contains the
experimental proof of a large gap between the filled and conduction bands.
However, it is difficult to conclude whether in equilibrium the conduction
gap is empty or not. Below we suggest an alternative interpretation for Ref. 
\cite{nature1} based not on the analogy with the Mott transition, but on the
analogy with the quantum crystals (QC).

There is a strong similarity between the ultracold particles in optical
lattices and atoms in QC, such as solid helium, in which the tunneling is
sufficiently high to ensure band motion of atoms unless prohibited by large
on-site repulsion (see review \cite{and1} and references therein). In helium
crystals, the atomic band motion is banned, as for all MI, when all the
lattice sites are occupied by \emph{identical} particles with the occupancy
equal to one. If, however, some of the lattice sites are empty, nothing
prohibits tunneling of atoms from the occupied onto the vacant sites
leading, as a result of translational symmetry, to the band motion of
vacancies, \emph{i.e., }to the formation of peculiar band quasiparticles -
vacancy waves. Similar quasiparticles are formed when some of the atoms
occupy interstitial sites and can tunnel through the QC. A slightly
different situation occurs when some of the lattice sites are occupied by
atoms of a different kind from the host matrix (impurities). The impurities
can also tunnel through the QC despite the fact that each site still has the
occupancy equal to one. The tunneling constant for impurities is smaller
than for the vacancies since the exchange of places between the impurity and
host atoms involves high-energy intermediate states with either double
on-site occupancy or the atom in an interstitial position. The impurity-host
exchanges could be so low that a more efficient mechanism of impurity motion
could be the vacancy-assisted diffusion. The behavior of vacancy and
impurity waves in QC is well understood \cite{and1}. However, the most
exciting possibility in QC - superfluidity and BEC in the system of vacancy
waves - has not been realized for "classical" QC, namely, solid $^{4}He$
despite two decades of intensive efforts (Refs. \cite{good2,good1} and
references therein). The reason is that there are no zero-temperature
vacancies in solid $^{4}He$: with lowering temperature, the concentration of
vacancies drops exponentially always remaining insufficient for BEC.

The alkali atoms in optical lattices resemble QC with a very appealing
difference: the BEC in the system of vacancy or "impurity" waves could be
within reach. When the occupancy of the individual wells is close to an
integer, the system resembles QC with either a small concentration of
vacancies or "excessive" atoms. If the occupancy is slightly below the
integer $K$, $K=1+\mathrm{Int}\left[ n/N\right] ,$ where $n$ and $N$\ are
the densities of atoms and lattice sites, then the density of vacancies is $%
n_{v}=KN-n\ll N$. If the occupancy slightly exceeds the integer number, then
the density of "excessive" atoms $n_{e}\equiv N-n_{v}$ is $n_{e}=n-N\mathrm{%
Int}\left[ n/N\right] \ll N$. Since the tunneling probabilities $t$ are the
same for vacancies and "excessive" atoms, $t_{v}=t_{e}$ (in both cases, an
atom tunnels to an empty site), many properties of the system are symmetric
with respect to the vacancies and excessive atoms. [This is not so for usual
QC in which the lattice potential is built of the atom interaction and the
potential relief is different for a vacancy and an "excessive" atom].

Below we consider the situation with large on-site interaction $U$ when the
lattice system with the integer site occupation would become a MI making the
BEC transition impossible. The analogy with QC automatically excludes fully
occupied lowest on-site states and does not require a concept of
counter-superfluidity \cite{kuk1}.

In the tight binding approximation for vacancies in a simple cubic lattice, 
\begin{equation}
\epsilon _{v}\left( \mathbf{p}\right) =\Delta /2-2t_{v}\sum \cos \left(
p_{i}a/\hbar \right)   \label{spec1}
\end{equation}%
where $\Delta =12t_{v}$ is the bandwidth. At large $U$, the fixed chemical
potential $\mu _{v}$ is finite in contrast to $\mu _{v}=0$ for
thermally-activated vacancies in solid helium. When $n_{v}$ is small, the
BEC transition temperature for the vacancies can be determined using the
standard equations for lattice gases with low band filling:%
\begin{equation}
T_{c}=6.6a^{2}t_{v}n_{v}^{2/3}  \label{e1}
\end{equation}%
and similar for excessive atoms. A good extrapolation between these two
limiting cases is 
\begin{equation}
T_{c}=6.6a^{4}t_{v}n_{v}^{2/3}\left( N-n_{v}\right) ^{2/3}.  \label{e2}
\end{equation}%
With the above values of $E_{r}$ and $a$, the estimate for the BEC
transition in the vacancy liquid (VL) is $T_{c}\sim 3\times 10^{-7}\left(
t_{v}/E_{r}\right) x_{v}^{2/3}\left( 1-x_{v}\right) ^{2/3}$ $\mathrm{K}$
where $x_{v}=a^{3}n_{v}$ is the fraction of unoccupied highest on-site
states.

Though $T_{c}$ for the VL can be quite high, there are two reasons why $T_{c}
$ is lower than the BEC temperature for a free gas. First, the density of
participating particles is lower (only the vacancies or the excessive atoms
in the highest on-site state are subject to the condensation). In experiment 
\cite{nature1} with the occupancy between 2 and 3, this leads, at least, to
a factor $5^{-2/3}$ in $T_{c}$\ and even stronger lowering of $T_{c}$ if the
system is close to an integer occupancy. Second, the effective mass of
vacancy waves or excessive particles, $m^{\ast }=\hbar ^{2}/2ta^{2}=m\left(
E_{r}/\pi ^{2}t\right) $, could be much larger than the mass of the free
atoms $m$. In experiment, one has limited control over the vacancy
concentration. On the other hand, the tunneling frequency depends
exponentially on the intensity of the laser beams. This makes $m^{\ast }$ a
readily adjustable parameter that can make the superfluid transition in the
VL $\left( \ref{e2}\right) $ observable.

All this suggests a probable alternative to the Mott transition for
experiment \cite{nature1}. At large $t$, the experiment confirmed the
presence of BEC, probably in the "free" alkali gas rather than in the VL. At
small $t$, the experiment showed the absence of the condensate. However, the
experiment, by design, cannot distinguish between the MI and the VL. It is
very likely that the experiment showed the superfluid - VL transition rather
than the superfluid - MI transition.

The analogy with QC allows one make several other predictions. First, the
role of "impurity waves" can be played by atoms in the different hyperfine
states. Experimentally, such impurities can be studied by means similar to
the NMR methods for $^{3}He$ diffusion in solid $^{4}He$. In principle,
impurities become band particles spread across the system with the tunneling
frequency $t_{i}$. However, when the on-site interaction $U$ is large, $%
t_{i} $ for direct exchange of impurities with the host atoms and,
therefore, the impurity wave bandwidth become very small, of the order of $%
t_{i}\sim t_{v}^{2}/U\ll t_{v}$. When $U$ is sufficiently large, such direct
tunneling can be often disregarded.

If the number of upper-state vacancies is noticeable, the vacancy-assisted
processes dominate the impurity motion with an effective intersite tunneling
rate $t_{i}\sim t_{v}x_{v}\gg t_{v}^{2}/U$ (in this context, the asymmetry
of the vacancy-assisted motion \cite{kesh1} is not important). At $T>T_{c}$,
the vacancy-assisted processes are responsible for independent tunneling
transitions between the adjacent sites. This is not a band motion but a more
traditional impurity diffusion with an effective diffusion coefficient 
\begin{equation}
D_{i}\sim \frac{t_{v}a^{2}}{\hbar }\frac{n_{v}}{N-n_{v}}.  \label{d1}
\end{equation}%
For the vacancy-assisted impurity tunneling, in contrast to the pure system (%
\ref{e2}), $t_{i}$ and the mean free paths are not symmetric with respect to 
$n_{v}\rightarrow 0$ and $n_{v}\rightarrow N$. In the former limit, the free
paths are atomic while in the latter limit the impurities recover their band
properties with the large mean free path determined by the impurity-impurity
scattering or the scattering by the few remaining upper-state host atoms.

The situation changes dramatically after the vacancy system undergoes the
superfluid transition (\ref{e2}). Then the impurity becomes a completely
delocalized quasiparticle in the vacancy superfluid background similar to $%
^{3}He$ impurities in the superfluid $^{4}He$ \cite{mey1}. The effective
mass of such quasiparticles at $T=0$ is $m_{i}^{\ast }\sim \hbar
^{2}/2t_{v}x_{v}$ and goes up with temperature with a decrease in density of
the vacancy condensate. The interaction effects in this quasiparticle gas
are negligible and the properties of the system can be evaluated using the
standard equations for an ideal lattice gas of quasiparticles. At low enough
temperatures, this impurity component of density $n_{i}$ will also undergo
its own BEC with%
\begin{equation}
T_{ci}\simeq 6.62t_{v}a^{5}n_{v}n_{i}^{2/3}=a^{3}n_{v}^{1/3}n_{i}^{2/3}T_{c}
\label{d3}
\end{equation}%
where $T_{c}$\ is the temperature for the vacancy condensation $\left( \ref%
{e1}\right) $. The emerging two-condensate system should exhibit properties
similar to those of liquid $^{3}He-\,^{4}He$ mixtures with two condensates
below the $^{3}He$ transition \cite{mey2}. Since this BEC in is based on the
vacancy-assisted tunneling, this two-condensate system is different from the
one considered in Ref. \cite{kuk1}.

This picture of the vacancy-assisted impurity motion works well when the
concentration of the hyperfine impurities $x_{i}=a^{3}n_{i}$ is low. At
higher $x_{i}$ the vacancy motion in this translationally inhomogeneous
environment is accompanied by the host-impurity permutations suppressing the
band motion. This is similar to the vacancy motion in solid $^{3}He$ with
disordered spin system. Then the vacancies autolocalize within the
homogeneous domains of the size%
\begin{equation}
R=\left[ \frac{\pi \hbar ^{2}}{2m^{\ast }NT\left[ \left( x_{i}-1\right) \ln
\left( 1-x_{i}\right) -x_{i}\ln x_{i}\right] }\right] ^{1/5}  \label{r1}
\end{equation}%
which are filled by particles in one hyperfine state (Nagaoka polarons). If
the density of vacancies is large, $n_{v}^{1/3}R\gtrsim 1$, this should lead
to the decomposition of the system into macroscopic hyperfine domains. The
difference between this decomposition and the vacancy-driven spin
polarization of solid $^{3}He$ \cite{mey3} is that the transition takes
place when the concentration of the zero-point vacancies and "polarization"
(the concentration of hyperfine components) are fixed. In contrast to the
formation of dynamic, transient domains in experiment \cite{jila1}, this
decomposition leads to stationary domains. If the hyperfine impurities are
bosons, this decomposition is not always necessary below the vacancy BEC.

One feature of the optical lattices is quite different from more
"traditional" QC such as helium. Since the periodic potential in QC is built
of atomic interaction, the vacancy motion is tied to the deformation of the
lattice. For the alkali atoms in optical lattices, the lattice is the
external potential of the laser beams and the particle displacement is
decoupled from the deformation of the lattice. As a result, the
low-frequency collective modes above and below BEC are decoupled from the
lattice variables.

An important issue for QC is the sensitivity of the narrow-band particles to
external fields. Since the energy of band particles cannot change by more
than the bandwidth $\Delta $, the external field $\Omega \left( \mathbf{r}%
\right) $ makes the motion finite and localizes the particles in the area of
the size $\delta r\sim a\sigma ,$ $\sigma =\Delta /\left( \partial \Omega
/\partial r\right) a$. For a usual QC, the important fields are the particle
interaction, lattice deformation, and external forces. In the optical
lattice, the most important field is the trapping potential $\Omega =\frac{1%
}{2}\alpha r^{2}$. In wide traps (small $\alpha $), this trapping potential
does not cause noticeable Umklapp processes and the overall Hamiltonian of a
particle (or a vacancy) in the optical lattice, $H=\epsilon \left( \mathbf{p}%
\right) +\frac{1}{2}\alpha r^{2}$, can be treated quasiclassically. Analysis
of this Hamiltonian can be performed in the momentum representation in which 
$\frac{1}{2}\alpha r^{2}=\frac{1}{2}\hbar ^{2}\alpha \nabla _{\mathbf{p}}^{2}
$ and the problem reduces to that for a particle with "mass" $1/\alpha $ in
the "potential" $\epsilon \left( \mathbf{p}\right) $. The quantum problem is
the simplest near the band minima where the quantized motion is harmonic
with the characteristic frequency $\omega _{0}^{\ast }=\left( 2t\alpha
\right) ^{1/2}a/\hbar $. Since the effective (tunneling) mass $m^{\ast
}=\hbar ^{2}/2ta^{2}$ is larger than the mass of free particles, this
frequency is $\left( m/m^{\ast }\right) ^{1/2}$ times lower than for the
free particles. The quantization in the trapping potential $\Omega \left( 
\mathbf{r}\right) $ is usually not important and the motion is close to
classical. At large\ $\sigma $\ the motion is unrestricted. When $\sigma
\rightarrow 1$, even the classical motion becomes compressed towards the
multiwell shells around the center of the trap with $\sigma $ giving the
thickness of the shell, to which the particle motion is restricted, in terms
of the well size $a$. In experiment \cite{nature1}, $\omega _{0}^{\ast }\sim
75\left( t/E_{r}\right) ^{1/2}$ $\mathrm{Hz}$ and is small while $\sigma
\gtrsim 100\left( t/E_{r}\right) $. The shells narrow to a single well layer
at the beam intensity for which $t/E_{r}\lesssim 0.01$.

The inhomogeneity of the trap also leads to the non-uniform spatial
redistribution of particles \cite{mott1}. If the change of the trapping
potential from well to well is large in comparison with temperature ($\Delta
/T\sigma \gg 1$), the shells with the lower energy are fully filled, have
integer population, and become the MI. The rest of the shells (most likely,
but not necessarily, the outer ones) will have the non-integer population
and resemble the VL with a rather large density of vacancies. In experiment 
\cite{nature1}, parameter $\Delta /T\sigma \sim 10^{-10}/T$ (with $T$ in 
\textrm{K}) seems to be small meaning insignificant redistribution of
particles between the shells. Even if this parameter were large, the system
would represent a thick shell with $\sigma \gtrsim 100\left( t/E_{r}\right) $
of coupled well layers in the quasi-2D VL state with the rest of the shells
in the MI state with filled upper levels. This may actually increase the BEC
temperature for the VL since the VL, though restricted to a lower number of
shells, can have a higher density of vacancies.

In summary, we explored the analogy between the alkali gases in optical
lattices with non-integer occupation and large on-site interaction with QC.
This analogy provides an alternative explanation for experiment \cite%
{nature1} as a transition between the BEC and VL states. The BEC transition
for the VL is predicted. The transition temperature seems to be within
experimental reach. The presence of a large number of unoccupied states
provides a vacancy-assisted mechanism for diffusion of hyperfine impurities
and can, sometimes, lead to a spatial decomposition of the system into pure
hyperfine components. The properties of the hyperfine mixture depend on
whether the system is above or below BEC temperature for the VL. At even
lower temperature one can observe the transition to the state with two -
vacancy and impurity - condensates which is different from Ref. \cite{kuk1}.

I am grateful to the referee for pointing out the importance of
redistribution of particles in traps. The work was supported by NSF grant
DMR-0077266.


\begin{thebibliography}{99}
\bibitem{cor1} E. A. Cornell and C. E. Wieman, Rev. Mod. Phys. \textbf{74},
875 (2002)

\bibitem{review2} A. J. Leggett, Rev. Mod. Phys. \textbf{73}, 307 (2001); F.
Dalfovo, S. Giorgini, L. P. Pitaevskii, and S. Stringari, Rev. Mod. Phys. 
\textbf{71}, 463 (1999)

\bibitem{jila1} H. J. Lewandowski, D. M. Harber, D. L. Whitaker, and E. A.
Cornell, Phys.Rev.Lett. \textbf{88}, 070403 (2002); M. \"{O}. Oktel and L.
S.\ Levitov, Phys.Rev.Lett. \textbf{88}, 230403 (2002); J. N. Fuchs, D. M.
Gangardt, and F. Lalo\"{e}, Phys.Rev.Lett. \textbf{88}, 230404 (2002); J. E.
Williams, T. Nikuni, and C. W. Clark, Phys.Rev.Lett. \textbf{88}, 230405
(2002); A. Kuklov and A. E. Meyerovich, Phys. Rev. A \textbf{66}, 023607
(2002)

\bibitem{lat1} G. Raithel et al., Phys. Rev. Lett. \textbf{78}, 630 (1997);
T. M\"{u}ller-Seydlitz et al., ibid., 1038 (1997); S. E. Hamann et al.,
ibid. \textbf{80}, 4149 (1998); L. Guidoni et al., ibid, \textbf{79}, 3363
(1997); K. I. Petsas, A. B. Coates, and G. Grynberg, Phys. Rev. A \textbf{50}%
, 5173 (1994); I. H. Deutsch and P. S. Jessen, ibid, \textbf{57}, 1972 (1998)

\bibitem{mott1} D. Jaksch, C. Bruder, J. I. Cirac, C. W. Gardiner, and P.
Zoller, Phys. Rev. Lett. \textbf{81}, 3108 (1998)

\bibitem{nature1} M. Greiner, O. Mandel, T. W. Hansch, and I. Bloch, Nature 
\textbf{415}, 39 (2002)

\bibitem{and1} A. F. Andreev, Defects and Surface Phenomena in Quantum
Crystals, in: \textit{Quantum Theory of Solids, }\ ed. I. M. Lifshits (Mir
Publishers, Moscow, 1982) pp. 11 - 69

\bibitem{good2} P. Remeijer, S. C. Steel, R. Jochemsen, G. Frossati, and J.
M. Goodkind, Sov.Phys. - Low Temp. Phys., \textbf{23}, 438 (1997) [Fiz.
Nizk. Temp. \textbf{23}, 586 (1997)]

\bibitem{good1} J. M. Goodkind, Phys. Rev. Lett. \textbf{89}, 095301 (2002)

\bibitem{kuk1} A.B. Kuklov and B.V. Svistunov, cond-mat/0205069 (2002)

\bibitem{kesh1} O. A. Andreeva, K. O. Keshishev, and A. D. Savischev,
Physica B: Cond. Matt. \textbf{284-288}, 343 (2000)

\bibitem{mey1} E.P. Bashkin, and A.E. Meyerovich, Adv. Phys. \textbf{30}, 1
- 92 (1981)

\bibitem{mey2} A.E.Meyerovich, in: \textit{Prog.Low Temp.Phys., }ed. by D.
F. Brewer (North-Holland, Amsterdam, \textit{1987), }Vol. 11, pp. 1-73;
A.E.Meyerovich, \textit{Spin-Polarized Phases of }$^{3}He$, in: Helium
Three, eds. W.P.Halperin and L.P.Pitaevski (Elsevier, Amsterdam) 1990, pp.
757 - 879

\bibitem{mey3} A. F. Andreev, V. I. Marchenko, and A. E. Meyerovich, JETP
Lett. \textbf{26}, 36 (1977)
\end{thebibliography}
\end{document}